\documentclass[aip,jmp]{revtex4-1}
\usepackage{graphics}
\usepackage{color}
\usepackage{amsmath}
\usepackage{amsthm} 
\usepackage{amssymb}
\usepackage{graphicx}
\usepackage{amsfonts}
\usepackage{times}
\usepackage{bbm}

\DeclareMathOperator{\tr}{tr}

\newtheorem{theorem}{Theorem}
\newtheorem{corollary}{Corollary}
\newcommand{\nn}{{\mathbbm{N}}}
\newcommand{\rr}{{\mathbbm{R}}}

\newcommand{\id}{{\mathbbm{1}}}
\newcommand{\me}{\mathrm{e}}
\newcommand{\mi}{\mathrm{i}}

\newcommand{\be}{\begin{equation}}
\newcommand{\ee}{\end{equation}}

\begin{document}


\title{\hspace{0.5cm}Dynamical error bounds for continuum discretisation via \\
\hspace{0.5cm}Gauss quadrature rules, -- a Lieb-Robinson bound approach}

\author{M.P. Woods}
\affiliation{University College of London, Department of Physics \& Astronomy, London WC1E 6BT,
United Kingdom}
\affiliation{Centre for Quantum Technologies, National University of Singapore, Singapore} 
\affiliation{QuTech, Delft University of Technology, Lorentzweg 1, 2611 CJ Delft, Netherlands}
\author{M.B. Plenio}
\affiliation{Institute f\"ur Theoretische Physik, Universit{\"a}t Ulm, D-89069 Ulm, Germany}


%

\date{Received: date / Accepted: date}
%
%

%

%

%
\begin{abstract}
Instances of discrete quantum systems coupled to a continuum of oscillators are ubiquitous
in physics. Often the continua are approximated by a discreate set of modes.
We derive error bounds on expectation values of system observables that have been time evolved
under such discretised Hamiltonians. These bounds  take on the form of a function of time
and the number of discrete modes, where the discrete modes are chosen according to Gauss
quadrature rules. The derivation makes use of tools from the field of Lieb-Robinson bounds
and the theory of orthonormal polynominals.
\end{abstract}
\maketitle

Instances of discrete quantum systems coupled to continua are ubiquitous in physics as
they describe open quantum systems, i.e. well-characterised systems under the control
of the experimenter that are in contact with a much larger and typically uncontrolled
environment \cite{RivasH2012}. Examples can be found in quantum optics \cite{knight15},
solid state and condensed matter physics \cite{Weiss2001} and recently quantum biology
\cite{mart13,mohseni2014quantum} to name just a few.
In numerical studies environments with continuous spectra are often modeled by a discrete
spectrum while in analytical work the reverse, i.e. replacing a discrete environmental
spectrum by a continuous one in a "continuum limit" \cite{Weiss2001}, is often convenient.
%
%
There have been many suggestions about how to best approximate continuous spectra by discrete
spectra for the evaluation of dynamical quantities and numerical studies into their efficiency.
These studies were, to the best of our knowledge, initiated by Rice in 1929 \cite{rice29_1,rice29_2,rice29_3}
in which the continuum was discretised to form a point spectrum with support at equally spaced
points. Later, it was suggested by Burkey and Cantrell  \cite{burk84} that a different
choice of discretisation would lead to a more accurate description of the dynamics. This later
idea was based on the fact that approximating integrals by discrete sums using Gauss quadrature
rules is often more efficient than the trapezoidal rule. A  bibliographical review of the subject
can be found in \cite{shore83}.
%
While sometimes estimates to some of the committed errors are given \cite{burk84,iyer11},
exact bounds for the quantity
\begin{equation}\label{aprox 0}
    \bigl|\tr\bigl[\hat{O}\me^{-\mi t\hat{H}_\textup{con}}\hat{\varrho}_0\me^{\mi t\hat{H}_\textup{con}}\bigr]  -\tr\bigl[\hat{O}\me^{-\mi t\hat{H}_\textup{dis}}\hat{\varrho}_0\me^{\mi t\hat{H}_\textup{dis}}\bigr]\bigr|,
\end{equation}
where $\hat O$ is an observable, $\hat{\varrho}_0$ is any initial state of potential interest,
$\hat{H}_\textup{con}$ is a Hamiltonian with absolutely continuous spectra and $\hat{H}_\textup{dis}$ is a Hamiltonian with pure
point spectra, do not appear to exist in the literature. Bounds of this form are of particular
relevance because they concern precisely the quantities of physical interest -- the expectation
values of local observables. In this article, we will derive bounds on the quantity Eq.
(\ref{aprox 0}) for physically relevant unbounded Hamiltonians, for discretisation schemes based
on Gauss quadrature rules.

In a different vein of research, in the context of lattice quantum systems, a bound introduced
by Lieb and Robinson \cite{lieb72} provides a measure for the speed of propagation of signals in a
spatially extended spin quantum system with finite range interaction, and bounds the decay of the
magnitude of signals propagating faster than this speed. The bound has been an indispensable tool
to prove many intriguing properties of many-body Hamiltonians and their ground and thermal states,
such as, e.g., the Lieb-Schulz-Mattis theorem \cite{LR_applications1,LR_applications2}, exponential
clustering of correlations \cite{brunoRober06,LR_applications3}, area laws for entanglement \cite{LR_applications4,RevEiserMarcusMartin2010}, efficient approximations of ground and thermal
states and dynamics \cite{osbo06,LR_applications6,LR_applications7}, and speed limits on the distribution
of correlations and entanglement \cite{LR_applications8,LR_applications9,LR_applications10}. Naturally,
since the original statement of the bound in 1972, there have been many refinements and generalizations,
see, e.g., \cite{LR_refinements1,LR_refinements2,LR_refinements3,Nachtergaele2008,Marcus_bosonic_LR,LR_refinements4} and references therein.

More precisely, the Lieb-Robinson bound states that there is a velocity $v>0$ and constants $\mu>0$
and $K>0$, which depend on the details of the lattice and the Hamiltonian, such that the operator
norm $\|\cdot\|$ of the commutator of two local observables $\hat{A}$ and $\hat{B}$, separated
by a distance $|x|$, is bounded by
\begin{equation}
    \label{(1.1)}
    \|[e^{\mi \hat Ht}\hat A e^{-\mi \hat Ht},\hat B]\|\leq K\|\hat A\|\|\hat B\|e^{-\mu(|x|-vt)}.
\end{equation}
Often, the development of a new Lieb-Robinson bound goes hand in hand 
with the development of new physical theorems. Here we show that the newly developed Lieb-Robinson 
bound in \cite{ourLR}, has an application in an area of research which has, up to now, only been 
probed numerically. Namely, it will be one of two key ingredients in our proof of bounds for Eq. 
(\ref{aprox 0}). The other key ingredient, will be a unitary transformations based on the theory 
of orthogonal polynomials from a non-local Hamiltonian to a local one (an infinite lattice with 
local coupling), to which a Lieb-Robinson bound applies. These two ingredients will be combined to 
achieve the desired result. First, the non-local Hamiltonian will be written as an infinite lattice 
with local interaction via a unitary operation. Secondly, this lattice will be spatially truncated  
and the Lieb-Robinson bound will be applied to estimate the error involved in the truncation. Finally, 
a unitary transformation will be applied to the truncated lattice to write it in the form of the 
desired discretised (non-local) Hamiltonian. When viewed from the perspective of the discretised 
non-local Hamiltonian, the Lieb-Robinson distance $|x|$ (described in Eq. \eqref{(1.1)}) will no 
longer play the role of a distance, but instead; will determine the number of pure point spectra 
sampled from the continuum.

\section{The Hamiltonian}\label{The Hamiltonian}
In this section we define the Hamiltonians we will be considering in this article\footnote{This
will be extended to include multiple copies of the bosonic bath Hamiltonian in section
\ref{Multiple baths extension}.}. We consider a quantum system coupled to a bosonic bath. The
Hilbert space $\mathcal{S}$ of the system carries a ``free'' Hamiltonian $\hat H_{\mathcal{S}}$,
while the bosonic bath is described by the Fock space $\mathcal{B}:=\mathrm{\Gamma}(\mathfrak{h})$ over
the mode Hilbert space $\mathfrak{h}$, with the free Hamiltonian $d\hat{\mathrm{\Gamma}}(G)$ arising from
the second quantization of the positive self-adjoint operator $G$ on $\mathfrak{h}$. The coupling 
is via one mode $h\in\mathfrak{h}$ and a bounded in operator norm self-adjoint system operator
$\hat A_\mathcal{S}$, so that altogether we have on $\mathcal{S}\otimes\mathcal{B}$ the Hamiltonian
\begin{equation}\label{the bathtub}
    \hat H=\hat H_{\mathcal{S}}\otimes\id_\mathcal{B}+\id_{\mathcal{S}}\otimes d\hat{\mathrm{\Gamma}}(G)+\hat A_\mathcal{S}\otimes\hat\Phi(h),
\end{equation}
where $\hat\Phi$ denotes the usual field operator. $G$ is a function $g$ of momentum variables
$k$ (known as the \textit{dispersion relation}), and can be expressed in terms of bath creation and
annihilation operators as
\be\label{eq:bath def}
    d\hat{\mathrm{\Gamma}}(G)=\int d{k}\ g({k})a_{{k}}^\dag a_{{k}}
\ee
and
\be\label{eq:int temr}
    \hat\Phi(h)=\int d{k}\ h({k})\bigl(a_{{k}}^\dag+a_{{k}}\bigr).
\ee
The form Eq. (\ref{the bathtub}) is often referred to as a Pauli-Fierz Hamiltonian \cite{dere99},
specialized in our case by allowing one interaction term. We note that although $\hat A_\mathcal{S}$
is bounded, there are no constraints on the system Hamiltonian $\hat H_\mathcal{S}$, it can be
unbounded or otherwise and comprise of bosons, fermions, spins etc. Definining the \textit{free} Hamiltonian as $\hat H_0=\hat H_{\mathcal{S}}\otimes\id_\mathcal{B}+\id_{\mathcal{S}}\otimes d\hat{\mathrm{\Gamma}}(G) $, Eq. (\ref{the bathtub}) is well
defined on $\mathcal{D}(\hat H)=\mathcal{D}(\hat H_0)$ if \cite{Reed1975}
\be
\int\frac{h^2(k)}{g(k)}dk<\infty.
\ee
It includes the spin-boson-Model achieved by letting $\mathcal{S}=\mathbb{C}^2$, $\hat H_\mathcal{S}=\mathcal{\alpha}\hat\sigma_z$, and $\hat A_\mathcal{S}=\hat\sigma_x$, where
$\mathcal{\alpha}$ is a positive constant and $\hat\sigma_x,$ $\hat\sigma_z$ are the Pauli
matrices.
Let $E(dx)$ be the projection valued spectral measure of $G$. Then we form the scalar measure
\begin{equation}\label{mu}
    \mu_0(d\omega)=\langle h,E(d\omega)h\rangle\ .
\end{equation}
The measure is absolutely continuous with respect to the Lebesgue measure, and we can write
$\mu_0(dx)= \pi^{-1} J(x)dx$ with the \textit{spectral density}\footnote{here the factor of
$1/\pi$ is for conformity with \cite{Weiss2001}} $J$. 
When $g$ is monotone, it is defined as
\begin{equation}\label{SD_eq}
    J(\omega)=\pi h\left(g^{-1}(\omega)\right)^2\ \left|\frac{dg^{-1}(\omega)}{d\omega}\right|,
\end{equation}
where $g^{-1}$ is the inverse of $g$. If $g$ is not monotone, we would additionally have a sum
over inverse images $g^{-1}(\{\omega\}),$ and when the momentum variable has more dimensions, we
would also have an integral over the inverse image. The minimally closed interval containing its
support is $[\omega_{min},\omega_{max}]$ with $\omega_{min}:=\textup{inf}\; g\geq 0,$ $\omega_{max}:=\textup{sup}\; g$.
The case $\omega_{min}=0$ is called \textit{massless} where as $\omega_{min}>0$ is known as \textit{massive}.
We will be dealing with the case $\omega_{max}<\infty$, hence $\mu_0(dx)$ is determinate with all
moments finite (see e.g. \cite{chiara}). For later purposes, it will be convenient to recall 
that Hamiltonian Eq. \eqref{the bathtub} is isomorphic to the so-called \textit{standard form} defined 
on page 5 of \cite{misc14}. We summarise this standard form here, for the  convenience of the reader.
We observe that the bath in Hamiltonian Eq. \eqref{the bathtub} is fully specified by the triple 
$(\mathfrak{h},G,h)$. Suppose now we have another environment system $(\tilde{\mathfrak{h}},\tilde G,\tilde h)$ 
and a unitary operator $U : \mathfrak{h} \rightarrow \tilde{\mathfrak{h}}$
such that $U h = \tilde h$ and $U G U^\dag = \tilde G$. Under such an isomorphism all the details of 
system-environment are mapped into each other. This is formally done by the unitary operator 
$\hat{\mathrm{\Gamma}} (U) : \mathrm{\Gamma}(\tilde{\mathfrak h}) \rightarrow \mathrm{\Gamma}(\tilde{\mathfrak h})$. 
Note that the scalar measure Eq. \eqref{mu} is invariant under this isomorphisum $U$. In fact, 
it completely determines the triple $(\mathfrak{h},G,h)$ up to an isomorphisum. Indeed, we are permitted to set
\begin{align}\label{eq:std form}
\begin{split}
\tilde{\mathfrak{h}}&=L^2(\rr^+,\mu),\\
(\tilde G\psi)(x)&=x \psi(x)\;\; \forall\, \psi\in \tilde{\mathfrak{h}},\\
\tilde{h}(x)&=1.
\end{split}
\end{align}
The unitary operator defining the isomorphism is given by $\left( U \me^{\mi t G}h\right)(x)=\me^{\mi t x}.$ The triple $(\tilde{\mathfrak{h}},\tilde G,\tilde h)$ defined by Eq. \eqref{eq:std form}, is referred to as the Standard form of $(\mathfrak{h}, G, h)$. We observe that when Eqs. \eqref{eq:bath def} and \eqref{eq:int temr} are written in standard form, $k$ is a scalar. We will use the standard form in theorems \ref{theorem dis 1} and \ref{theorem dis 2}.

\section{Discretisation of the continuous bath}\label{Discretisation of the continuous bath}
Let $P_n(x)$ be the real orthonormal polynomial of order $n$ with respect to the measure
$\mu_0(dx)=\pi^{-1}J(x)dx$ so that
\be
    \int \mu_0(dx) P_n(x)P_m(x)=\delta_{m,n},\quad n,m\in \nn_0.
\ee
The existence and uniqueness (up to a real phase) of these orthonormal polynomials is well
established \cite{gaut10}. We can use them to define a discretised Hamiltonian
\be\label{discrete 1}
    \hat H_L=\hat H_\mathcal{S}\otimes\id_\mathcal{B}+\hat A_\mathcal{S}\otimes\sum_{n=1}^L h_n^{(L)}(c_n^\dagger+c_n)+\id_\mathcal{S}\otimes\sum_{n=1}^L\omega_n^{(L)}c_n^\dagger c_n,
    \quad L\in\nn_+
\ee
where $\omega_n^{(L)}$ are the zeros of $P_L(x)$ and
\be
    h_n^{(L)}=\frac{1}{\sqrt{\sum_{k=0}^{L-1} P^2_k(\omega_n^{(L)})}}.
\ee
The $c_n$ $(c_m^\dagger)$ on $\mathcal{B}$ are bosonic annihilation (creation) operators (and can
be expressed in terms of the field operators of Hamiltonian Eq. (\ref{the bathtub}) as detailed
by Eq. \eqref{eq:c n} in the proof). The discretised Hamiltonian, Eq. (\ref{discrete 1}) has a
nice interpretation in terms of Gauss quadrature rules: $\{\omega_n^{(L)} \}_{n=1}^L$ are the
Gauss knots and $\{(h^{(L)}_n)^2\}_{n=1}^L$ are the Gauss weights for the weight function $J(\cdot)/\pi$
(see section 1.4.2 for an introduction to Gauss quadrature and last paragraph of section 1.4.1
and Eq. (3.1.7) for the nodes and weights \cite{gaut10}). What is more, the theory of orthogonal
polynomials has established that $\omega_{min}\leq \omega_n^{(L)} \leq \omega_{max}$ for all
$L,n$ and $\{\omega_n^{(L)}\}_{n=1}^{L}$ interlace $\{\omega_n^{(L+1)}\}_{n=1}^{L+1}:$
\begin{equation}\label{interlace 1}
    \omega_{L+1}^{(L+1)}<\omega_{L}^{(L)}<\omega_{L}^{(L+1)}<\omega_{L-1}^{(L)}<\ldots<\omega_{1}^{(L)}<\omega_{1}^{(L+1)},
\end{equation}
thus discretisation in terms of the zeros of orthogonal polynomials corresponds to a natural
way of discretising a continuum. We note that the spectral density of Hamiltonian Eq. (\ref{discrete 1})
forms a pure point measure. This is in stark contrast with the spectral density of Hamiltonian
Eq. (\ref{the bathtub}), which is the weight function of an absolutely continuous measure with
respect to the Lebesgue measure.

This discretisation will lead to errors in time evolved observables on the system degrees
of freedom that can be bounded as determined by the following theorem. For self-adjoint
operators of the form $\hat O=\hat O_\mathcal{S}\otimes\id_\mathcal{B}$, $\hat O_\mathcal{S}\in\mathcal{S}$
and initial normalised quantum states $\hat\varrho_0$ on $\mathcal{S}\otimes\mathcal{B}$ we find the following result.
\begin{theorem}\label{theorem dis 1}
    For any system observable $\hat O,$ $\| \hat O\|<\infty$ the error introduced on its expectation value at any time $t\geq 0$ when it's dynamics are approximated by the discretised Hamiltonian $\hat H_L,$ is bounded by
    \begin{align}\label{error H_L}
        \begin{split}
        &\bigl|\tr\bigl[\hat{O}\me^{-\mi t\hat{H}}\hat{\varrho}_0\me^{\mi t\hat{H}}\bigr]-\tr\bigl[\hat{O}\me^{-\mi t\hat{H}_L}\hat{\varrho}_0\me^{\mi t\hat{H}_L}\bigr]\bigr|^2 \\
        &\leq 8\eta_0 \|\hat O\|^2 \frac{\|\hat A_\mathcal{S}\|}{\omega_{max}}\frac{(\omega_{max} t)^{L+1} }{(L+1)!}  \left( \me^{\omega_{max} t}+1\right)\left( \|\vec
        \gamma_0\|^{1/2}+\eta_0\|\hat A_\mathcal{S}\|t \right),
     \end{split}
    \end{align}
for $L=1,2,3,\ldots$ where $\eta_0=\sqrt{\frac{2}{\pi}\int dx J(x)}$, and the operator $\vec \gamma_0$
encodes information regarding 2-point correlation functions in $\mathcal{B}$ of the initial state. More
precisely,
\be\label{eq:gamma mat thorem 1}
\vec \gamma_0=
\left(\begin{array}{cc}
\vec \gamma_{xx} & \vec \gamma_{xp}\\
\vec \gamma_{px} & \vec \gamma_{pp}
\end{array}\right),
\ee

where $\left[\vec\gamma_{xx}\right]_{n,m}=\tr[\hat x_n \hat x_m\hat\varrho_0]$, $\left[\vec\gamma_{xp}\right]_{n,m}=\tr[\hat x_n \hat p_m\hat\varrho_0]$, $\left[\vec\gamma_{px}\right]_{n,m}=\tr[\hat p_n \hat x_m\hat\varrho_0]$, $\left[\vec\gamma_{pp}\right]_{n,m}=\tr[\hat p_n \hat p_m\hat\varrho_0]$, $n,m\in\nn_+$ with
\begin{align}
    \begin{split}\label{eq:x p expansion}
       \hat x_n&=\frac{1}{\sqrt{2}}\int \mu_0^{(1/2)}(dx) P_{n-1}(x) \left(a_x^\dag+a_x\right),\\
       \hat p_n&=\frac{\mi}{\sqrt{2}} \int \mu_0^{(1/2)}(dx) P_{n-1}(x) \left(a_x^\dag-a_x\right), \quad n\in \nn_+
    \end{split}
\end{align}
on $\mathcal{B}$ where $\mu_0^{(1/2)}(dx):=\sqrt{J(x)/\pi}\,dx$. Here $\hat x_n$ and $\hat p_n$ have been written in standard form (see section \ref{The Hamiltonian}).
\end{theorem}
See section \ref{LR bunds proof} for the proof of this theorem and appendix \ref{sec:alternative expressions for gamma 0} for alternative expressions for $\|\vec \gamma_0\|$ and examples including when $\tr_\mathcal{S}[\hat \varrho_0]$ is the vacuum state.

Theorem \ref{theorem dis 1} provides, to the best of our knowledge, for the first time, a bound on the
error that results from approximating a bath with absolutely continuous spectrum by a bath with pure point
spectrum in the form of a sum over discrete modes and vice versa.
The Lieb-Robinson light cone is achieved by choosing a tangent surface to the r.h.s. of Eq. \eqref{error H_L} bounding this non-linear function from above.
Prior to this work discretisations of continuum baths had a long history of being probed numerically
(see \cite{burk84,Kazansky1997,Shen08,iyer11,Ines15} and references here in). A fundamental insight was provided
by Burkey and Cantrell when they numerically observed that using the evenly spaced knots specified by the
trapezoid rule (which is referred to as \textit{Rice discretization}) seems not to be the most efficient
way to discretise a continuum \cite{burk84}. As pointed out in \cite{Shen08}, their choice of Gauss quadrature
rules to discretise the continuum is for the weight function $J(\cdot)/\pi,$ and thus theorem \ref{theorem dis 1}
applies to it. We will thus refer to the particular choice of Gauss weights and knots in Eq. \eqref{discrete 1}
as \textit{Burkey-Cantrell} discretisation.

However, Eq. (\ref{discrete 1}) is not the unique way to use Gauss quadrature rules to discretise
the continuum. The next theorem will present a sharper bound compared to that of Eq. \eqref{error H_L}
by discretising the bath according to the weight function $J(\sqrt{\cdot})/\pi$ instead. Via a trivial
change of variable followed by using the properties of $\mu_0(dx)$, one verifies that the measure
\be\label{eq:phonon measure}
    \mu_1(dx):=\pi^{-1}J(\sqrt{x})dx,
\ee
is determinate, with all moments finite. Again, we define $P_n'(x)$ as the real orthonormal
polynomial of order $n$ with respect to the measure $\mu_1(dx)$:
\be\label{eq:ortho poly 2nd measure}
    \int \mu_1(dx) P_n'(x)P_m'(x)=\delta_{m,n},\quad n,m\in \nn_0.
\ee
We use these orthogonal polynomials to define a discretised Hamiltonian:
\be\label{discrete 2}
    \hat H_L'=\hat H_\mathcal{S}\otimes\id_\mathcal{B}+\hat A_\mathcal{S}\otimes\sum_{n=1}^L \frac{h_n^{\prime(L)}}{\sqrt{2\bar{\omega}_n^{(L)}}}(c_n^{\prime\dagger}+c_n')+\id_\mathcal{S}
    \otimes\sum_{n=1}^L\bar{\omega}_n^{(L)} c_n^{\prime\dagger} c_n', \quad L\in \nn_+
\ee
where $\bar{\omega}_n^{(L)}=\sqrt{\omega_n^{\prime(L)}},$ $\omega_n^{\prime(L)}$ are the zeros of
$P_L'(x)$ and
\be
    h_n^{\prime(L)}=\frac{1}{\sqrt{\sum_{k=0}^{L-1} P^{\prime2}_k(\omega_n^{\prime(L)})}}.
\ee
$c'_i$ $(c_j^{\prime\dagger})$ on $\mathcal{B}$ are bosonic annihilation (creation) operators (and
can be expressed in terms of the field operators of Hamiltonian Eq. (\ref{the bathtub}) as detailed
by Eq. \eqref{eq:d n} in the proof that is found in section 8.).

Here $\{\omega_n^{\prime(L)} \}_{n=1}^L$ are the Gauss knots and $\{(h^{\prime(L)}_n)^2\}_{n=1}^L$
are the Gauss weights for the weight function $J(\sqrt{\cdot})/\pi$. The knots satisfy
$\omega_{min}^2\leq \omega_n^{\prime(L)}\leq \omega_{max}^2$ for all $L,n$ and satisfy the same
interlacing properties as the Gauss knots $\{\omega_n^{(L)} \}_{n=1}^L$ observed in Eq. (\ref{interlace 1}).
To the best of our knowledge, this is a new discretisation, which, unlike the Burkey-Cantrell
discretisation, has not been probed numerically. We now state analogous theorems to that of
Eq. (\ref{error H_L}), but with a smaller r.h.s. for otherwise unchanged parameters. We will
make the distinction between when Hamiltonian Eq. (\ref{the bathtub}) is massless and massive.
\begin{theorem}\label{theorem dis 2}
    For any system observable $\hat O,$ $\| \hat O\|<\infty$ the error introduced on it's expectation value at any time $t\geq 0$ when it's dynamics are approximated by the discretised Hamiltonian $\hat H_L'$, is bounded by
    \begin{itemize}
    \item [] \textup{1)} Massive case \textup{($\omega_{min}>0$)}\\
    \begin{align}\label{error H_L'}
        \begin{split}
        &\bigl|\tr\bigl[\hat{O}\me^{-\mi t\hat{H}}\hat{\varrho}_0\me^{\mi t\hat{H}}\bigr]-\tr\bigl[\hat{O}\me^{-\mi t\hat{H}_L'}\hat{\varrho}_0\me^{\mi t\hat{H}_L'}\bigr]\bigr|^2 \\
        &\leq D_1\frac{(\omega_{max} t)^{2L+1} }{(2L+1)!}  \left( \me^{\omega_{max} t}+1\right)\left( \|\vec \gamma_0'\|^{1/2}+\eta_1\|\hat A_\mathcal{S}\| t \right),
        \end{split}
    \end{align}
    \item [] \textup{2)} Massless case \textup{($\omega_{min}=0$)}\\
    \begin{align}\label{error H_L'2}
        \begin{split}
        &\bigl|\tr\bigl[\hat{O}\me^{-\mi t\hat{H}}\hat{\varrho}_0\me^{\mi t\hat{H}}\bigr]
        -\tr\bigl[\hat{O}\me^{-\mi t\hat{H}_L'}\hat{\varrho}_0\me^{\mi t\hat{H}_L'}\bigr]\bigr|^2 \\
        &\leq D_1 \frac{(\omega_{max} t)^{2L+1}}{(2L+1)!}\left( \me^{ \omega_{max} t}+1\right) \left(
        \|\vec \gamma_0'\|^{1/2}+\eta_1\|\hat A_\mathcal{S}\|\frac{\me^{\omega_{max}t}-1}{\omega_{max}}
        \right) \me^{\omega_{max} t}
        \end{split}
    \end{align}
\end{itemize}
for $L=1,2,3,\ldots$ where $D_1=4\eta_1\|\hat O\|^2 \frac{\|\hat A_\mathcal{S}\|}{\omega_{max}}$,
$\eta_1=\sqrt{\frac{1}{\pi \omega_{max}}\int dx J(\sqrt{x})}$, and the operator $\vec \gamma_0'$ encodes information regarding 2-point correlation functions in $\mathcal{B}$ of the initial state. More precisely,
\be\label{eq:gamma mat thorem 2}
    \vec \gamma_0'=
    \left(\begin{array}{cc}
    \vec \gamma_{xx}' & \vec \gamma_{xp}'\\
    \vec \gamma_{px}' & \vec \gamma_{pp}'
    \end{array}\right),
\ee
where $\left[\vec\gamma_{xx}'\right]_{n,m}=\tr[\hat x_n' \hat x_m'\hat\varrho_0]$, $\left[\vec\gamma_{xp}'\right]_{n,m}=\tr[\hat x_n' \hat p_m'\hat\varrho_0]$,
$\left[\vec\gamma_{px}'\right]_{n,m}=\tr[\hat p_n' \hat x_m'\hat\varrho_0]$,
$\left[\vec\gamma_{pp}'\right]_{n,m}=\tr[\hat p_n' \hat p_m'\hat\varrho_0]$,
$n,m\in\nn_+$ with
\begin{align}
    \begin{split}\label{eq:x p expansion 2}
    \hat x_n'&=\sqrt{\omega_{max}}\int \mu_0^{(1/2)}(dx) P'_{n-1}(x^2)
    \left(a_x^\dag+a_x\right),\\
    \hat p_n'&=\frac{\mi}{\sqrt{\omega_{max}}} \int \mu_0^{(1/2)}(dx) P'_{n-1}(x^2)\,x
    \left(a_x^\dag-a_x\right), \quad n\in \nn_+
    \end{split}
\end{align}
on $\mathcal{B}$. Here $\hat x_n'$ and $\hat p_n'$ have been written in standard form (see section \ref{The Hamiltonian}).
\end{theorem}
See section \ref{LR bunds proof} for proof. See section \ref{sec:Alternative expressions for vec gamma_0'}
for alternative expressions for $\|\vec \gamma_0'\|$\\

{\em Observations} The bound that has been obtained in theorem \ref{theorem dis 2} up to constant factors achieves
the same error as the Burkey-Cantrell discretisation but with only half the number of knot points.
This observation suggests an improved discretisation method using Gauss quadrature rules and may also
find applications in analytical and numerical work.

Interestingly, for theorems (\ref{theorem dis 1}) and (\ref{theorem dis 2}) we used a Lieb-Robinson bound to derive a relation between two non-local Hamitonians (in Eq. (\ref{the bathtub}) and its discretised counterparts Eqs. (\ref{discrete 1}) and (\ref{discrete 2}), every harmonic oscillator is coupled \textit{directly} to the system via $\hat A_\mathcal{S}$). But now, in these non-local Hamiltonians, the notion of a distance that is  normally associated with a lattice in the Lieb-Robinson bound plays the role of the number of Gauss knot points $L$.

Note also that for $\omega_{max}=\infty$, the upper bound in Eqs. (\ref{error H_L}) and (\ref{error H_L'})
diverge. Under the minimal assumptions that have been made in this manuscript regarding the initial
state $\hat \varrho_0$, this is to be expected since for $\omega_{max}=\infty$ we are sampling an
unbounded interval with a finite number of sample points\footnote{This observation, gives an intuitive
explanation to why certain types of Hamiltonians can be proven \textit{not} to admit a Lieb-Robinson
bound \cite{eise09}, since when the lattice is diagonalised via a unitary transformation, one should
expect to observe similar divergences}. The theory of orthogonal polynomials also tells us that if the
support of the spectral density is gapped, i.e. the spectral density vanishes strictly in an interval
$[\omega_i,\omega_f]$ with $\omega_{min}<\omega_i<\omega_f<\omega_{max}$, a situation known to occur in physical systems such as photonic crystals \cite{prior13},
then the discretised bath will have at most one discrete mode in the gap $[\omega_i,\omega_f]$.

\section{Multiple baths extension}\label{Multiple baths extension}
Often in physical settings, the quantum system on $\mathcal{S}$ is coupled to multiple continuous
baths via different interaction terms. In such circumstances, the Hamiltonian is on $\mathcal{S}\otimes
\mathcal{B}^{\otimes N}$ and reads
\be
    \label{the bathtub multi}
    \hat H_\text{mul}=\hat H_\mathcal{S}\otimes\id_\mathcal{B}^{\otimes N} + \sum_{m=1}^N \hat H^{(m)},
    \quad N\in \nn_+
\ee
where
\begin{align}
    \begin{split}
        \hat H^{(m)}=&\hat A_\mathcal{S}^{(m)}\otimes\id_\mathcal{B}^{\otimes m-1}\otimes\left( \int dk h^{(m)}(k) (a_k^{(m)\dag}+a_k^{(m)}) \right)\otimes\id_\mathcal{B}^{\otimes N-m}\\
        &+\id_\mathcal{S}\otimes\id_\mathcal{B}^{\otimes m-1}\otimes\left(\int dk g^{(m)}(k)a^{(m)\dag}_k a_k^{(m)}\right)\otimes\id_\mathcal{B}^{\otimes N-m}.
    \end{split}
\end{align}
Here each individual bath and interaction term $\hat H^{(m)}$ is defined as in Eq. (\ref{the bathtub})
and will have its own spectral density $J^{(m)}.$ Since they are independent bosonic baths, we also
have $[a_x^{(n)},a_y^{(m)\dag}]=\delta_{n,m}\delta(x-y),$ $[a_x^{(n)},a_y^{(m)}]=0.$ In analogy with
Eqs. (\ref{discrete 1}) and (\ref{discrete 2}), we can define discretised versions of $\hat H^{(m)}$
according to the two discretisation schemes considered in this article. Similarly to Eq.
(\ref{discrete 1}), for Burkey-Cantrell discretisation, we define
\begin{align}\label{discrete 1 n}
    \begin{split}
        \hat H^{(0,m)}_L=&\hat A_\mathcal{S}^{(m)}\otimes\id_\mathcal{B}^{\otimes m-1}\otimes\left(\sum_{n=1}^L h_n^{(m,L)}(c_n^{(m)\dag}+c_n^{(m)})\right)\otimes\id_\mathcal{B}^{\otimes N-m}\\
        &+\id_\mathcal{S}\otimes\id_\mathcal{B}^{\otimes m-1}\otimes\left(\sum_{n=1}^L\omega_n^{(m,L)}c_n^{(m)\dag} c_n^{(m)}\right)\otimes\id_\mathcal{B}^{\otimes N-m}, \quad L,m\in \nn_+
    \end{split}
\end{align}
where the Gauss knots $\omega_n^{(m,L)}$ and Gauss weights $(h_n^{(m,L)})^2$ are calculated
from the weight function $J^{(m)}(\cdot)/\pi,$ for which we denote its minimum and maximum frequencies
by $\omega_{min}^{(m)}$ and $\omega_{max}^{(m)}$ respectively (see paragraph below Eq. (\ref{SD_eq})).
Similarly to the second discretisation procedure we considered, Eq. (\ref{discrete 2}), we define
\begin{align}\label{discrete 2 n}
    \begin{split}
        \hat H_L^{(1,m)}=&\hat A_\mathcal{S}^{(m)}\otimes\id_\mathcal{B}^{\otimes m-1}\otimes\left(\sum_{n=1}^L \frac{h_n^{\prime(m,L)}}{\sqrt{2\bar{\omega}_n^{(m,L)}}}(c_n^{\prime(m)\dagger}+c_n^{\prime(m)})\right)
        \otimes\id_\mathcal{B}^{\otimes N-m}\\
        &+\id_\mathcal{S}\otimes\id_\mathcal{B}^{\otimes m-1}\otimes\left(\sum_{n=1}^L\bar{\omega}_n^{(m,L)} c_n^{\prime(m)\dagger} c_n^{\prime(m)}\right) \otimes\id_\mathcal{B}^{\otimes N-m}, \quad L,m\in \nn_+.
    \end{split}
\end{align}
We can now define the discretised Hamiltonian
\be
    \hat H^{\vec q}_{\vec L}=\hat H_\mathcal{S}\otimes\id_\mathcal{B}^{\otimes N}
    + \sum_{m=1}^N \hat H_{L_m}^{(q_m,m)}, \quad N\in \nn_+
\ee
where $\vec q=(q_1,q_2,\ldots,q_N)$ is a binary string and $\vec L=(L_1,L_2,\ldots,L_N).$
With the definitions
\begin{align}
    \Big(\Delta^{(0,m)}_{L}(t)\Big)^2 = D_0^{(m)}\frac{(\omega_{max}^{(m)} t)^{L+1} }{(L+1)!}
    \left( \me^{\omega_{max}^{(m)} t}+1\right)\left( \|\vec \gamma_0^{(m)}\|^{1/2} +
    \eta_0^{(m)}\|\hat A_\mathcal{S}\|t \right),
\end{align}
\begin{align}
    \begin{split}
        \Big(\Delta^{(1,m)}_{L}(t)\Big)^2&=
        \begin{cases}
            f_L^{(m)}(t)  &\mbox{if } \omega_{min}^{(m)}>0\\
        	g_L^{(m)}(t)  & \mbox{if } \omega_{min}^{(m)}=0,
        \end{cases}\\
        f_L^{(m)}(t)&=C_1^{(m)} \frac{(\omega_{max}^{(m)} t)^{2L+1} }{(2L+1)!} ( \me^{ \omega_{max}^{(m)} t}+1)
        \left( \|\vec\gamma_0^{(m)}\|^{1/2}+\eta_1^{(m)}\|\hat A_\mathcal{S}^{(m)}\|t \right)\\
        g_L^{(m)}(t)&=C_1^{(m)} \frac{(\omega_{max}^{(m)} t)^{2L+1}}{(2L+1)!}( \me^{ 2\omega_{max}^{(m)} t}+1)\left( \|\vec \gamma_0^{(m)}\|^{1/2}+\eta_1^{(m)}(\me^{\omega_{max}^{(m)}t}-1)\|\hat A_\mathcal{S}^{(m)}\| \right)\me^{\omega_{max}^{(m)} t}
    	\end{split}
	\end{align}	
	
\begin{align}
    \begin{split}
        D_0^{(m)}&=8\eta_0^{(m)} \|\hat O\|^2 \|\hat A_\mathcal{S}^{(m)}\|/\omega_{max}^{(m)},\quad
        \eta_0^{(m)}=\sqrt{\frac{2}{\pi}\int dx J^{(m)}(x)},\\
        C_1^{(m)}&=4 \eta_1^{(m)} \|\hat O\|^2 \|\hat A_\mathcal{S}^{(m)}\|/\omega_{max}^{(m)},\quad \eta_1^{(m)}=\sqrt{\frac{1}{\pi\omega_{max}^{(m)}}\int dx J^{(m)}(\sqrt{x})},
    \end{split}
\end{align}
where $\Delta^{(q,m)}_{L}\geq 0,$ we have the following bounds.
\begin{corollary}\label{multi cor}
    For any system observable $\hat O,$ $\| \hat O\|<\infty$ the error introduced on it's expectation value at any time $t\geq 0$ when it's dynamics are approximated by the discretised Hamiltonian $\hat H^{\vec q}_{\vec L}$, is bounded by
    \be\label{multi eq}
    \left| \tr\bigl[\hat{O}\me^{-\mi t\hat{H}_\textup{mul}}\hat{\varrho}_0\me^{\mi t\hat{H}_\textup{mul}}\bigr]-\tr\bigl[\hat{O}\me^{-\mi t \hat H^{\vec q}_{\vec L}}\hat{\varrho}_0\me^{\mi t\hat H^{\vec q}_{\vec L}}\bigr] \right|\leq \sum_{m=1}^N \Delta^{(q_m,m)}_{L_m}(t).
    \ee
\end{corollary}
See section \ref{multiple chains extension proof} for proof. We thus see that the error
incurred scales linearly in the number of discretised continua.

\section{Explicit examples}\label{Explicit examples}
For a range of spectral densities from the literature we will now present the explicit expressions
for the frequencies and Gauss weights.

Consider the semi-circle law spectral density:
\be
    J(\omega)=C\sqrt{(\omega_{max}-\omega)(\omega-\omega_{min})},
\ee
for some constant $C>0$. This is the weight function of the Chebyshev polynomials of the second
kind on the interval $[\omega_{min},\omega_{max}]$. Their corresponding orthogonal polynomials
and their zeros are known explicitly \cite{gaut10}. The zeros of the $n$th order Chebyshev
polynomial on $[-1,1]$ are
\be
    x_k=\cos \left( \frac{k}{n+1}\pi \right),\quad k=1,\ldots,n
\ee
and hence for the Burkey-Cantrell discretisation (Eq. (\ref{discrete 1})), the discrete frequencies are
\be
    \omega_k^{(L)}=\frac{(\omega_{min}-\omega_{max})}{2}\cos \left( \frac{k}{L+1}\pi \right) +\frac{(\omega_{max}+\omega_{min})}{2}.
\ee
For the spectral density of the \textit{Rubin model} \cite{Weiss2001} we find
\be
    J(\omega)=C\sqrt{(\omega^2_{max}-\omega^2)(\omega^2-\omega^2_{min})},
\ee
for some constant $C>0$. From Eq. (\ref{ph map dens}), we have that this corresponds to
a Chebyshev weight function of the second kind for the measure $J(\sqrt{\cdot})/\pi$ used
in Eq. \eqref{eq:ortho poly 2nd measure}. Thus we have for the 2nd discretisation method
(Eq. (\ref{discrete 2}))
\be
    \bar{\omega}_k^{(L)}=\sqrt{\frac{(\omega^2_{min}-\omega^2_{max})}{2}\cos \left( \frac{k}{L+1}\pi \right) +\frac{(\omega^2_{max}+\omega^2_{min})}{2}.}
\ee
Note that since the Chebyshev polynomials are known explicitly, we can also find explicitly the
$h_n^{(L)}$ coefficients in both the above examples.

Now we will consider the frequently considered power-law spectral densities
\be
    J(\omega)=2\pi \alpha (\omega_{max}-\omega_{min})(\omega-\omega_{min})^s,\quad -1<s
\ee
which include the sub-ohmic $s<1$, ohmic $s=1$ and the super-ohmic $s>1$ case. We can use
the Jacobi Polynomials to describe the measures $J(\cdot)/\pi$ and $J(\sqrt{\cdot})/\pi$.
For the Jacobi polynomials $P_n^{(\alpha,\beta)}(x)$ on $[-1,1]$, we have the Buell inequalities
for their zeros $x_i$ (\cite{gobo75}, pg 125):
\be
    \frac{i+(\alpha+\beta-1)/2}{n+(\alpha+\beta+1)/2}\pi<\nu_i<\frac{i}{n+(\alpha+\beta+1)/2}\pi,\quad i=1,\ldots,n
\ee
where $\nu_i=\arccos x_i$ and $-1/2\leq \alpha\leq 1/2$, $-1/2\leq \beta\leq 1/2$, excluding the case $\alpha^2=\beta^2=1/4$.
Hence for the Burkey-Cantrell discretisation (Eq. (\ref{discrete 1}))
\begin{align}\label{pa f J}
    \begin{split}
        &\left[ 1-\cos\left(\frac{k}{L+(s+1)/2}\pi\right)\right]\frac{\omega_{max}-\omega_{min}}{2}
        < \omega_k^{(L)}-\omega_{min} \\
        &<\left[1-\cos\left(\frac{k+(s-1)/2}{L+(s+1)/2}\pi\right)\right]
        \frac{\omega_{max}-\omega_{min}}{2}
    \end{split}
\end{align}
for $-1/2\leq s\leq 1/2$. For the massless ($\omega_{min}=0$) 2nd discretisation method (Eq. (\ref{discrete 2})),
\begin{align}\label{ph f J}
    \begin{split}
        &\omega_{max}\sqrt{\left[ 1-\cos\left(\frac{k}{L+(s/2+1)/2}\pi\right)\right]/2}< \bar{\omega}_k^{(L)}\\
        &<\omega_{max}\sqrt{\left[1-\cos\left(\frac{k+(s/2-1)/2}{L+(s/2+1)/2}\pi\right)\right]/2}
    \end{split}
\end{align}
for $-1< s\leq 1$. Note that other bounds for the zeros are known for the values of
$-1<s$ not covered here. See for example \cite{elbe94}. In all the examples in this
section, the corresponding orthogonal polynomials are known explicitly and thus one can achieve explicit expressions for the Gauss weights too.

\section{Conclusion}
We derive two error bounds for dynamical observables when discretising a
continuum of harmonic oscillators according to Gauss quadrature rules. For one of the bounds,
numerical studies have probed this discretisation numerically in the past. For the second case,
no prior numerical studies have been performed. The second bound achieves a sharper bound for the
same parameters, suggesting that the second Gauss quadrature discretisation method may be
more efficient.

To prove these results, we make use of a unitary mapping based on orthogonal polynomials 
and a Lieb-Robinson bound, providing yet another application to these powerful tools. Attempts 
to understand the errors endured by dynamical observables as a function of time when approximating 
a Hamiltonian with absolutely continuous spectrum with one of pure point spectrum, although 
receiving interest since 1929 by various authors, has up to now only consisted in numerical studies.

\textit{Acknowledgements -- } M.W. would like to thank Marcus Cramer for motivating discussions. He
also acknowledges support from MOE Tier 3A Grant MOE2012-T3-1. MBP ackowledges support
by an Alexander von Humboldt Professorship, the ERC grant BioQ and the EU projects PAPETS
and SIQS.

\section{Proofs}\label{Proofs}
\subsection{Proofs of theorems \ref{theorem dis 1} and \ref{theorem dis 2}}\label{LR bunds proof}
For convenience, we will prove both theorems in parallel. The proof proceeds along three main steps.
First, we will state unitary transformations of the bath modes which allow Eq. (\ref{the bathtub})
to be written as a Hamiltonian in which the system on $\mathcal{S}$ couples to the first
site of a nearest neighbour coupled harmonic chain. This chain is then truncated at finite length
and a Lieb-Robinson bond is deployed to achieve error bounds for the expectation
value of systems operators. Subsequently, the finite chain is transformed back to a non-local Hamiltonian
and it is demonstrated that this achieves the discretised Hamiltonians as formulated
in theorems \ref{theorem dis 1} and \ref{theorem dis 2}.

In \cite{misc14}, it was shown that if $J(\cdot)$ has all moments finite, the Hamiltonian Eq. (\ref{the bathtub})
can be written in the forms
\begin{align}\label{marticle mapping hamiltonian eq}
\begin{split}
\hat H=& \hat H_\mathcal{S}\otimes\id_\mathcal{B}+\sqrt{\beta_0(0)} \hat A_\mathcal{S}\otimes(b_0(0)+b_0^\dagger(0))\\
&+\id_\mathcal{S}\otimes\sum_{n=0}^\infty \alpha_n(0)b_n^\dagger(0) b_n(0) +\sqrt{\beta_{n+1}(0)}(b_{n+1}^\dagger(0) b_n(0) +h.c.)
\end{split}
\end{align}
and
\begin{align}\label{phonon chain hamiltonian}
\begin{split}
\hat H=& \hat H_{S}\otimes\id_\mathcal{B}+\sqrt{\frac{\beta_0(1)}{\omega_{max}}}\hat A_\mathcal{S}\otimes\hat X_0\\
&+\id_\mathcal{S}\otimes\sum_{n=0}^{\infty} \left(\frac{\sqrt{\beta_{n+1}(1)}}{\omega_{max}}\hat X_n\hat X_{n+1}+ \frac{\alpha_n(1)}{2\omega_{max}}\hat X^2_n +\frac{\omega_{max}}{2}\hat P_n^2\right)\;\\
&+K\id_\mathcal{S}\otimes\id_\mathcal{B}.
\end{split}
\end{align}
where
\be\label{eq:phnon ops}
\hat X_n:=\sqrt{\omega_{max}}\,(b_n^\dag(1)+b_n(1)),\quad \hat P_n:=\mi (b_n^\dag(1)-b_n(1))/(2\sqrt{\omega_{max}}),\quad n\in\nn_0,
\ee
are position and momentum operators satisfying the canonical commutation relations. It was established in Theorem 33 of \cite{misc14} that the operators $b_n(q),b^\dag_n(q)$ are well defined on $\mathcal{D}(\mathfrak{n}^{1/2})$, where $\mathfrak{n}=d\hat{\mathrm{\Gamma}}(\id_\mathfrak{h})$ is the second quantised boson number operator. $K\in \rr$ thus w.l.o.g.
from here on it will be set to zero since it does not contribute to Eq. (\ref{error H_L'}).
Now we will state a recently derived locality bound \cite{ourLR}.

The Hamiltonian the bound applies to is system with Hamiltonian $\hat{H}_\mathcal{S}$ on $\mathcal{S}$ coupled to a semi-infinite nearest neighbour bosonic chain of the form
\begin{equation}\label{the bath}
\hat{H}_\mathcal{B}=\frac{1}{2}\sum_{i,j=0}^\infty\left(\hat{x}_iX_{i,j}\hat{x}_j+\hat{p}_iP_{i,j}\hat{p}_j\right),
\end{equation}
where $X_{i,j}=X_{j,i}\in\rr$,  $P_{i,j}=P_{j,i}\in\rr$ where $X_{i,j}=P_{i,j}=0$ for $|i-j|>1$.
The system-bath coupling is of the form
$\hat{h}\otimes\hat{x}_0,$ where $\hat{h}$ acts on the system and is bounded in operator norm.
The total Hamiltonian reads
\begin{equation}\label{tot ham}
\hat{H}=\hat{H}_\mathcal{S}\otimes\id_\mathcal{B}+\hat{h}\otimes\hat{x}_0+\id_\mathcal{S}\otimes\hat{H}_\mathcal{B}.
\end{equation}
We then define a spatially truncated Hamiltonian
\begin{equation}
\begin{split}
\label{eq:H_L}
\hat{H}_L&=\hat{H}_\mathcal{S}\otimes\id_\mathcal{B}+\hat{h}\otimes\hat{x}_0+\id_\mathcal{S}\otimes\hat{H}^L_\mathcal{B},\\
\hat{H}^L_\mathcal{B}&=\frac{1}{2}\sum_{i,j=0}^{L-1}\left(\hat{x}_iX_{i,j}\hat{x}_j+\hat{p}_iP_{i,j}\hat{p}_j\right)
\end{split}
\end{equation}
and the constant $c$ such that $\|X_LP_L\|^{1/2}\le c$. Theorem 1 in \cite{ourLR} for $X,P>0$ or $X=P$ gives us the bound
\begin{equation}\label{bound 1}
\frac{\bigl|\tr\bigl[\hat{O}\me^{-\mi\hat{H}t}\hat{\varrho}_0\me^{\mi\hat{H}t}\bigr]-\tr\bigl[\hat{O}\me^{-\mi\hat{H}_Lt}\hat{\varrho}_0\me^{\mi\hat{H}_Lt}\bigr]\bigr|^2}{4\|\hat{O}\|^2\|\hat{h}\|/c}
\le
\frac{C(\|\vec \gamma_0\|^{\frac{1}{2}}+\|\hat{h}\|t)(c t)^{L+1}(\me^{ct}+1)}{(L+1)!}
,
\end{equation}
where
\begin{equation}
\begin{split}
C&= \|P_L\||X_{L-1,L}|/c^2+|P_{L-1,L}|/c
\end{split}
\end{equation}
and
\begin{equation}\label{eq:gamma 0 general}
\vec \gamma_0=\left(\begin{array}{cc}
\vec\gamma_{xx} & \vec\gamma_{xp}\\
\vec\gamma_{px} & \vec\gamma_{pp}
\end{array}\right),\;\;\; [\vec \gamma_{ab}]_{i,j}=\tr[\hat{a}_i\hat{b}_j\hat{\varrho}_0],\;\;\; a,b=x,p,
\end{equation}
collects the two-point bath correlations in the initial state. Also note that $|X_{L-1,L}|\le \|X_L\|\leq \|X\|$ and $|P_{L-1,L}|\le \|P_L\|\leq \|P\|$.
If $X,P>0$ or $X=P$ are not satisfied, we use theorem 3 in \cite{ourLR} to achieve the bound. Let $c'$ such that $\{\|X\|,\|P\|\}\leq c',$ then for all $X,$ $P$
\begin{equation}
\begin{split}\label{bound2}
&\frac{\bigl|\tr\bigl[\hat{O}\me^{-\mi\hat{H}t}\hat{\varrho}_0\me^{\mi\hat{H}t}\bigr]-\tr\bigl[\hat{O}\me^{-\mi\hat{H}_Lt}\hat{\varrho}_0\me^{\mi\hat{H}_Lt}\bigr]\bigr|^2}{4\|\hat{O}\|^2\|\hat{h}\|/c}\\
&\hspace{4cm}\le
\frac{C\me^{c^\prime t}(\|\vec \gamma_0\|^{1/2}+\|\hat{h}\|
\frac{\me^{c^\prime t}-1}{c^\prime})(ct)^{L+1}(\me^{ct}+1)}{(L+1)!}.
\end{split}
\end{equation}
If $P\propto\id$, we may replace the factor $(ct)^{L+1}/(L+1)!$ by $(ct)^{2L+1}/(2L+1)!$ in the R.H.S. of Eqs. \eqref{bound 1}, \eqref{bound2}. We can readily apply these bounds to Eqs. (\ref{marticle mapping hamiltonian eq}) and (\ref{phonon chain hamiltonian}). First define position and momentum operators
\be\label{eq:particle pos momentum ops def}
\hat x_n(0):=(b^\dagger_n(0)+b_n(0))/\sqrt{2},\quad \hat p_n(0):=\mi(b^\dagger_n(0)-b_n(0))/\sqrt{2},\quad n\in\nn_0.
\ee
Comparing Eq. (\ref{the bath}) with Eqs. (\ref{marticle mapping hamiltonian eq}), (\ref{phonon chain hamiltonian}) and the definition of the Jacobi matrices $\mathcal{J}(d\lambda^q)$ (see Eq. (162) in \cite{misc14}), we find\footnote{Here we use the notation of \cite{misc14} for reference purposes. In this article, we have used the notation $\mu_0(dx)\equiv d\lambda^0(x)$ and $\mu_1(dx)\equiv d\lambda^1(x)$. We will continue to use the notation of \cite{misc14} throughout this proof for these two measures}\\
\begin{itemize}
\item[] For Eq. (\ref{marticle mapping hamiltonian eq}):
\be
X=P=\mathcal{J}(d\lambda^0),\quad \hat h=\sqrt{2\beta_0(0)}\hat A_\mathcal{S},\quad d\lambda^0(x)=\pi^{-1}J(x)dx.
\ee
\item[] For Eq. (\ref{phonon chain hamiltonian}):
\be
X=\frac{\mathcal{J}(d\lambda^1)}{\omega_{max}},\quad P=\id\;\omega_{max}, \quad\hat h=\sqrt{\frac{\beta_0(1)}{\omega_{max}}}\hat A_\mathcal{S},\quad d\lambda^1(x)=\pi^{-1}J(\sqrt{x})dx.\label{ph map dens}
\ee
\end{itemize}
From Eqs (15,156,160) in \cite{misc14},
\be
\beta_0(0)=\int dxJ(x)/\pi,\quad \beta_0(1)=\int dx J(\sqrt{x})/\pi.
\ee
Since the spectrum of a Jacobi matrix is equal to its minimally closed support interval \cite{Teschl2000}, we have for the Eqs. (\ref{marticle mapping hamiltonian eq}), (\ref{phonon chain hamiltonian}): $\|X\|=\|P\|=\sqrt{\|XP\|}=\omega_{max}$, and $X>0$ iff $\omega_{min}>0$. 
The r.h.s. of Eqs. (\ref{error H_L}), (\ref{error H_L'}) and (\ref{error H_L'2}) are a direct consequence of these bounds. We will now proceed to apply another unitary transformation. This time, we will apply it to the above spatially truncated Hamiltonians to write them in terms of Gauss quadrature.\\
Both Eqs. (\ref{marticle mapping hamiltonian eq}) and (\ref{phonon chain hamiltonian}) can be written in the compact form (see Eq. (162) in \cite{misc14})
\begin{align}\label{the generalised mapping in terms of the jacobi mat eq}
\begin{split}
H=&\hat H_{S}\otimes\id_\mathcal{B}+\sqrt{\beta_0(q)}\,\hat A_\mathcal{S}\otimes(b^\dagger(q)+b(q))\\
&+\id_\mathcal{S}\otimes\frac{q}{2}\left[ \vec b^\top (q)\left(\mathcal{J}(d\lambda^q)-\frac{q}{4}\id\right)\vec b(q) +h.c.\right]\\
&+\id_\mathcal{S}\otimes\vec b^\dagger(q) \left(\mathcal{J}(d\lambda^q)+\frac{q}{4}\id\right)\vec b(q),
\end{split}
\end{align}
where
\begin{align}
\begin{split}
\vec b^\dagger(q)&:=(b_0^\dagger(q),b_{1}^\dagger(q),b_{2}^\dagger(q),b_{3}^\dagger(q),\ldots),\\
\vec b(q)&:=(b_0(q),b_{1}(q),b_{2}(q),b_{3}(q),\ldots)^\top ,
\end{split}
\end{align}
and Eq. (\ref{marticle mapping hamiltonian eq}) is given when $q=0$ while Eq. (\ref{phonon chain hamiltonian})
when $q=1$. In the case of the truncated chains $\hat H_L$and $\hat H_L'$, we simply replace
$\mathcal{J}$ with $\mathcal{J}_{L}:=\mathcal{J}_{[1:L;1:L]},$ $\id$ with $\id_{[1:L;1:L]}$
and $\vec b$ with $\vec b_{[1:L]}$.
Since Jacobi matrices are real symmetric, they are diagonalisable via an orthogonal transformation. Thus
\begin{equation}
    \mathcal{J}_L(d\lambda^q)=O_{L}(d\lambda^q)D^{(L)}(d\lambda^q)O^{\top}_{L}(d\lambda^q)
\end{equation}
where $\textup{diag}D^{(L)}(d\lambda^q)=(\omega^{(L)}_1 (d\lambda^q),\ldots,\omega^{(L)}_L(d\lambda^q))$
are the eigenvalues of $\mathcal J (d\lambda^q)$ and $O_{L}(d\lambda^q)=(\vec v_1(d\lambda^q),\ldots,\vec v_L(d\lambda^q))$ with $\vec v_i(d\lambda^q)$ the normalised eigenvector for $\omega^{(L)}_i (d\lambda^q)$.
Let $\tilde P_n(d\lambda^q,x)$ be the real orthogonal polynomial of order $n$ corresponding to the
measure $d\lambda^q(x)$:
\be
    \int d\lambda^q(x) \tilde P_n(d\lambda^q;x)\tilde P_m(d\lambda^q;x)=||\tilde P_n ||_{d\lambda^q}^2\,\delta_{m,n},\quad n,m\in \nn_0
\ee
where $m=n$ defines $|| \tilde P_n ||_{d\lambda^q}$.
The set $\{\tilde P_n(x;d\lambda^q)\}_{n=0}^\infty$ exists and for a specific choice of
normalisation $\{|| \tilde P_n ||_{d\lambda^q}\}_{n=0}^\infty$, it is unique up to a real
phase \cite{gaut10}. It is known that  $\{\omega_i^{(j)}(d\lambda^q)\in \rr\}_{i=1}^j$ are
the zeros of $\tilde P_j(d\lambda^q;x)$ and
\begin{align}
\begin{split}
    &\vec v_i(d\lambda^q)=\\ &\vec 
    (\tilde P_0(d\lambda^q,\omega_i^{(L)}(d\lambda^q)),\ldots,\tilde  P_{L-1}(d\lambda^q,\omega_i^{(L)}(d\lambda^q)))^\top \frac{1}{\sqrt{\sum_{k=0}^{L-1}\tilde{P}^2_k(d\lambda^q;\omega_i^{(L)}(d\lambda^q))}}
\end{split}
\end{align}  \cite{gaut10}. 
Let us define
\be
    h_j^{(L)}(d\lambda^q):=\frac{\sqrt{\beta_0(d\lambda^q)}\tilde P_0(d\lambda^q;\omega_j^{(L)}(d\lambda^q))}
    {\sqrt{\sum_{k=0}^{L-1}\tilde{P}^2_k(d\lambda^q;\omega_j^{(L)}(d\lambda^q))}},\quad j\in \nn_+.
\ee
Hence for Eq. (\ref{marticle mapping hamiltonian eq}):
\be
\hat H_L=\hat H_\mathcal{S}\otimes \id_\mathcal{B}+\hat A_\mathcal{S}\otimes\sum_{n=1}^L h_n^{(L)}(d\lambda^0)(c_n^\dagger+c_n)+\id_\mathcal{S}\otimes\sum_{n=1}^L\omega_n^{(L)}(d\lambda^0)c_n^\dagger c_n
\ee
where $\vec c:=O^{\top}_L(d\lambda^0)\vec b(0)_{[1:L]},\;$ $\vec c^\dag:=\vec b^\dag(0)_{[1:L]} O_L(d\lambda^0)$.
We can easily verify that $[c_i,c_j^\dagger]=\delta_{i,j},$ $[c_i,c_j]=0$ and are hence bosonic creation annihilation operators. If we use orthonormal polynomials (i.e. $||\tilde P_n||_{d\lambda^q}=1$), then the zeroth order polynomial
is $\tilde P_0(d\lambda^q;x)=1/\sqrt{\beta_0(q)}$ and hence we find the results stated in theorem \ref{theorem dis 1}.
For  Eq. (\ref{phonon chain hamiltonian}), we have
\be
\hat H_L=\hat H_\mathcal{S}\otimes \id_\mathcal{B}+\hat A_\mathcal{S}\otimes\sum_{n=1}^L\frac{h_n^{(L)}(d\lambda^1)}{\sqrt{2\bar{\omega}_n^{(L)}}}(c_n^{\prime\dagger}+c'_n)+\id_\mathcal{S}\otimes\sum_{n=1}^L\bar{\omega}_n^{(L)}c_n^{\prime\dagger} c'_n
\ee
where $\bar{\omega}^{(L)}_n:=\sqrt{\omega^{(L)}_n(d\lambda^1)},$
\begin{align}
\begin{split}
 c'_n:=&\sqrt{\frac{\bar{\omega}_n^{(L)}}{2}}\Bigl[\left(1+\frac{1}{2\bar{\omega}_n^{(L)}}\right)\left[O^{\top}_L(d\lambda^1)\vec b(1)_{[1:L]}\right]_n \\
&+\left(1-\frac{1}{2\bar{\omega}_n^{(L)}}\right)\left[\vec b^\dag(1)_{[1:L]} O_L(d\lambda^1)\right]_n\Bigr]\end{split}\\
\begin{split}
c^{\prime\dagger}_n:=&\sqrt{\frac{\bar{\omega}_n^{(L)}}{2}}\Bigl[\left(1+\frac{1}{2\bar{\omega}_n^{(L)}}\right)\left[\vec b^\dag(1)_{[1:L]} O_L(d\lambda^1)\right]_n \\
&+\left(1-\frac{1}{2\bar{\omega}_n^{(L)}}\right)\left[O^{\top}_L(d\lambda^1)\vec b(1)_{[1:L]}\right]_n\Bigr]
\end{split}
\end{align}
with $[c'_i,c_j^{\prime\dagger}]=\delta_{i,j},$ $[c'_i,c'_j]=0$ thus achieving the results of theorem \ref{theorem dis 2}.
From the theory of orthogonal polynomials \cite{gaut10}, it is also known that $\omega_n^{(L)}(d\lambda^q)$ are contained in the support interval of $d\lambda^q$ and that the zeros of $P_n(d\lambda^q;x)$ alternate with those of $P_{n+1}(d\lambda^q;x),$ that is
\begin{align}
\begin{split}
&\omega_{n+1}^{(n+1)}(d\lambda^q)<\omega_{n}^{(n)}(d\lambda^q)<\omega_{n}^{(n+1)}(d\lambda^q)<\omega_{n-1}^{(n)}(d\lambda^q)<\ldots \\
&<\omega_{1}^{(n)}(d\lambda^q)<\omega_{1}^{(n+1)}(d\lambda^q)
\end{split}
\end{align}
where $\omega_{i}^{(n+1)},$ and $\omega_{j}^{(n)}$ are ordered in descending order.

Using Eqs. (158) and (161) in \cite{misc14}, we can establish the relation between the field operators $a(f)=\int dx f(x)a_x,$ $a^\dag(f)=\int dx f(x)a_x^\dag$ in Eq. (\ref{the bathtub}) and the operators $c_n,c_n^\dag,$  $c'_n, c^{\prime\dag}_n$.
\begin{align}
\begin{split}
c_n&=\sum_{j=0}^{L-1} \frac{\tilde P_{j}(d\lambda^0;\omega^{(L)}_n(d\lambda^0))}{\sqrt{\sum_{k=0}^{L-1}\tilde P^2_k(d\lambda^0;\omega_n^{(L)}(d\lambda^0))}}\;a(\vec \gamma^0_n), \label{eq:c n}\\
c_n^\dag&=\sum_{j=0}^{L-1} \frac{\tilde P_{j}(d\lambda^0;\omega^{(L)}_n(d\lambda^0))}{\sqrt{\sum_{k=0}^{L-1}\tilde P^2_k(d\lambda^0;\omega_n^{(L)}(d\lambda^0))}}\;a^\dag(\vec \gamma^0_n),
\end{split}\\
\begin{split}
c'_n&=\sqrt{\frac{\bar \omega_n^{(L)}}{2}}\left( \sum_{j=0}^{L-1}\frac{\tilde P_j(d\lambda^1;\omega_n^{(L)}(d\lambda^1))}{\sqrt{\sum_{k=0}^{L-1}\tilde P^2_k(d\lambda^1;\omega_n^{(L)}(d\lambda^1))}}\left( a^\dag(f_j^n)+a(g_j^n)\right) \right),\\
c_n^{\prime\dag}&=\sqrt{\frac{\bar \omega_n^{(L)}}{2}}\left( \sum_{j=0}^{L-1}\frac{\tilde P_j(d\lambda^1;\omega_n^{(L)}(d\lambda^1))}{\sqrt{\sum_{k=0}^{L-1}\tilde P^2_k(d\lambda^1;\omega_n^{(L)}(d\lambda^1))}}\left( a(f_j^n)+a^\dag(g_j^n)\right) \right),\label{eq:d n}
\end{split}
\end{align}
where $f_j^n(x):=\gamma_j^1(x)(1+g(x)/\bar \omega_n^{(L)}),\,$ $
g_j^n(x):=\gamma_j^1(x)(1-g(x)/\bar \omega_n^{(L)}).$
If $x\in \rr$,
\be
\gamma_n^q(x):=P_n(d\lambda^q;g^{q+1}(x))\sqrt{\frac{J(g^{q+1}(x))}{\pi}\left| \frac{dg(x)}{dx} \right|},
\ee
and can be generalised for higher dimensions by writing Hamiltonian Eq. \eqref{the bathtub} in  \textit{standard form} (see section \ref{The Hamiltonian}). We now turn our attention to $\vec \gamma_0$. We start with deriving its value for the case of theorem \ref{theorem dis 1}. From Eqs. \eqref{eq:gamma 0 general}, \eqref{marticle mapping hamiltonian eq} and recalling the definitions \eqref{eq:particle pos momentum ops def}, we find\footnote{Here the index labeling has been shifted by one to start from 0. This is for consistency with the notation in \cite{misc14}.} $[\vec \gamma_{xx}]_{i,j}=\tr[\hat x_{i-1}(0)\hat x_{j-1}(0)\hat\varrho_0],$ $[\vec \gamma_{xp}]_{i,j}=\tr[\hat x_{i-1}(0)\hat p_{j-1}(0)\hat\varrho_0],$ $[\vec \gamma_{px}]_{i,j}=\tr[\hat p_{i-1}(0)\hat x_{j-1}(0)\hat\varrho_0],$ $[\vec \gamma_{pp}]_{i,j}=\tr[\hat p_{i-1}(0)\hat p_{j-1}(0)\hat\varrho_0],$ $i,j\in\nn_+.$
From Eq. (160-161) in \cite{misc14} we have that
\begin{align}
\begin{split}\label{eq:x p expansion proof}
\hat x_n(0)&=\int \mu_0^{(1/2)}(dx) P_{n}(x) (a_x^\dag+a_x),\\
\hat p_n(0)&=\mi \int \mu_0^{(1/2)}(dx) P_{n}(x) (a_x^\dag-a_x), \quad n\in \nn_0,
\end{split}
\end{align}
thus achieving Eqs. \eqref{eq:gamma mat thorem 1}, \eqref{eq:x p expansion} in theorem \ref{theorem dis 1}.
We achieve the expression for $\| \vec \gamma'_0\|$ in theorem \ref{theorem dis 2} similarly. Namely, from Eq. \eqref{eq:phnon ops} and Eq. (158) in \cite{misc14}, we find
\begin{align}
\begin{split}\label{eq:x p expansion proof 2}
\hat X_n'&=\sqrt{\omega_{max}}\int \mu_0^{(1/2)}(dx) P'_{n}(x^2) \left(a_x^\dag+a_x\right),\\
\hat P_n'&=\frac{\mi}{\sqrt{\omega_{max}}} \int \mu_0^{(1/2)}(dx) P'_{n}(x^2)\,x \left(a_x^\dag-a_x\right), \quad n\in \nn_0
\end{split}
\end{align}
thus after the definitions $\hat x_n':=\hat X_{n-1}$, $\hat p_n':=\hat P_{n-1}$, $n\in\nn_+$ we achieve Eqs. \eqref{eq:gamma mat thorem 2}, \eqref{eq:x p expansion 2} in theorem \ref{theorem dis 2}.
 \qed
\subsection{Multiple chains extension proof}\label{multiple chains extension proof}
Here we will prove corollary \ref{multi cor}.
Let
\be
G^{(L_1,L_2,\ldots,L_N)}= 
\tr\bigl[\hat{O}\me^{-\mi t \hat H^{\vec q}_{\vec L}}\hat{\varrho}_0\me^{\mi t\hat H^{\vec q}_{\vec L}}\bigr]
\ee
 where we will denote the scenario that the $nth$ bath has not been discretised, by replacing $L_n$ by $\infty$. We now add and subtract $G$ $N\!-\!1$ times to the r.h.s. of Eq. (\ref{multi eq}) each time discretising one more bath and starting with only one bath discretised. We then apply the triangle inequality and arrive at
\begin{align}
\begin{split}
&\left| \tr\bigl[\hat{O}\me^{-\mi t\hat{H}_\textup{mul}}\hat{\varrho}_0\me^{\mi t\hat{H}_\textup{mul}}\bigr]-\tr\bigl[\hat{O}\me^{-\mi t \hat H^{\vec q}_{\vec L}}\hat{\varrho}_0\me^{\mi t\hat H^{\vec q}_{\vec L}}\bigr] \right| \\ &\leq \left| 
G^{(\infty,\ldots,\infty,\infty)}
-G^{(\infty,\ldots,\infty,L_N)}  \right|
 +\left| G^{(\infty,\ldots,\infty,L_N)}-G^{(\infty,\ldots,\infty,L_{N-1},L_N)}  \right|\\ &\quad+\ldots+\left| G^{(\infty,L_2,L_3,\ldots,L_N)}-
G^{(L_1,L_2,L_3,\ldots,L_N)}
\right|.
\end{split}
\end{align}
In every pair on the r.h.s. of the inequality, there is always one bath which is discretised in one of the $G$ terms but is not discretised for the other $G$ term. We can thus define all the other baths to be part of the system Hamiltonian and then apply theorem \ref{theorem dis 1} or theorem \ref{theorem dis 2} to it. This gives us Eq. (\ref{multi eq}).\qed


\appendix

\section{Appendix: Alternative expressions for $\|\vec\gamma_0\|$, $\| \vec\gamma_0'\|$ and examples} In this appendix we derive alternative expressions for the constants $\|\vec \gamma_0\|$, $\|\vec \gamma_0'\|$ which appear in theorems \ref{theorem dis 1} and \ref{theorem dis 2} respectively. These alternative expressions come in the form of Eqs. \eqref{eq:op nr gamma functional form} and \eqref{eq:op nr gamma functional form 2} respectively. In addition, we  calculate explicitly $\|\vec \gamma_0\|$, $\|\vec \gamma_0'\|$ for some  examples cases of particular initial quantum states $\hat \varrho_0\in \mathcal{S}\otimes\mathcal{B};$ see sections \ref{sec:examples gamma}, and \ref{sec:examples gamma'} respectively. We start with $\|\vec \gamma_0\|$.
\subsection{Alternative expressions for $\|\vec\gamma_0\|$}\label{sec:alternative expressions for gamma 0}
Recall that the measure $\mu_0(dx)=\pi^{-1}J(x)dx$ is determinate, and thus $\{P_k\}_{k=0}^\infty$ form a complete orthonormal system in $L^2(\mu_0(dx))$ (see \cite{riesz23}, \cite{mario95}). Hence Eq. \ref{eq:x p expansion} represents an expansion of
\begin{align}
\begin{split}\label{eq:x p f}
\hat x(f):&=\frac{1}{\sqrt{2}}\int \mu_0^{(1/2)}(dx) f(x) \left(a_x^\dag+a_x\right),\\
\hat p(f):&=\frac{\mi}{\sqrt{2}} \int \mu_0^{(1/2)}(dx) f(x) \left(a_x^\dag-a_x\right),
\end{split}
\end{align}
where $f\in L^2(\mu_0(dx))$ and recall $\mu_0^{(1/2)}(dx):=\sqrt{J(x)/\pi}\,dx$. Namely,
\begin{align}
\begin{split}
\hat x(f)&=\sum_{n=1}^\infty c_n(f) \hat x_n, \\
\hat p(f)&=\sum_{n=1}^\infty c_n(f) \hat p_n,\quad c_n\in\rr.
\end{split}
\end{align}
\subsubsection{Basis invariance of $\|\vec \gamma_0\|$}\label{Basis invariance of vec gamma 0}
One can perform any unitary transformation of the operators $\{\hat x_n,\hat p_n\}_{n=1}^\infty$ into a new set $\{\hat{\underline{x}}_n,\hat{\underline{p}}_n\}_{n=1}^\infty$ with relationship $\vec R=\vec U \vec{\underline{R}}$, $\vec U^\dag \vec U=\vec U \vec U^\dag=\vec \id$, $\vec R:=(\hat x_1,\hat x_2,\ldots,\hat p_1,\hat p_2,\ldots)^\top,\,$ $\vec{\underline{R}}:=(\hat{\underline{x}}_1,\hat{\underline{x}}_2,\ldots,\hat{\underline{p}}_1,\hat{\underline{p}}_2,\ldots)^\top$. Similarly to before we can define
\be
\vec{\underline{\gamma}}_0=
\left(\begin{array}{cc}
\vec{\underline{\gamma}}_{xx} & \vec{\underline{\gamma}}_{xp}\\
\vec{\underline{\gamma}}_{px} & \vec{\underline{\gamma}}_{pp}
\end{array}\right),
\ee
where $\left[\vec{\underline{\gamma}}_{xx}\right]_{n,m}=\tr[\hat{\underline{x}}_n \hat{\underline{x}}_m\hat\varrho_0]$, $\left[\vec{\underline{\gamma}}_{xp}\right]_{n,m}=\tr[\hat{\underline{x}}_n \hat{\underline{p}}_m\hat\varrho_0]$, $\left[\vec{\underline{\gamma}}_{px}\right]_{n,m}=\tr[\hat{\underline{p}}_n \hat{\underline{x}}_m\hat\varrho_0]$, $\left[\vec{\underline{\gamma}}_{pp}\right]_{n,m}=\tr[\hat{\underline{p}}_n \hat{\underline{p}}_m\hat\varrho_0]$ for $n,m\in\nn_+$. It immediately follows by direct substitution $\vec \gamma_0= \vec U \vec{\underline{\gamma_0}} \vec U^\dag$ and thus
\be\label{eq:gamma norm equiv}
\|\vec \gamma_0\|=\|\vec{\underline{\gamma_0}}\|.
\ee
Eq. \eqref{eq:gamma norm equiv} shows that the two-point  correlation functions can be provided in any unitarily equivalent basis. Indeed, we now show that this feature allows us to write $\|\vec \gamma_0\|$ in a particularly appealing form.

\subsubsection{Alternative expression for $\|\vec \gamma_0\|$}\label{sec:When vec gamma_0 is self-adjoint}
Denoting the complex conjugation by $(\cdot)^*$, we have $(\tr[\hat x_n\hat p_m\hat\varrho_0])^*=\tr[(\hat x_n\hat p_m\hat\varrho_0)^\dag]=\tr[\hat p_m \hat x_n\hat \varrho_0]$ $\forall$ $n,m\in\nn_+$, where we have used the invariance of the trace under cyclic permutations and the self-adjointness of $\hat x_n$ , $\hat p_n$, $\hat \varrho_0$. We thus find
\be\label{eq:gamma self adjoint}
\vec \gamma_0^\dag=
\left(\begin{array}{cc}
\vec{\gamma}_{xx} & \vec{\gamma}_{xp}\\
\vec{\gamma}_{px} & \vec{\gamma}_{pp}
\end{array}\right)^\dag=
\left(\begin{array}{cc}
\vec{\gamma}_{xx}^\dag & \vec{\gamma}_{px}^\dag\\
\vec{\gamma}_{xp}^\dag & \vec{\gamma}_{pp}^\dag
\end{array}\right)
=\vec\gamma_0.
\ee
From Eq. \eqref{eq:gamma self adjoint} we see that $\vec \gamma_0$ is self-adjoint and hence its operator norm takes on the form
\be\label{eq:gamma self adjt}
\|\vec \gamma_0\|=\sup_{\|\vec v\|=1}\; \langle \vec \gamma_0 \vec v, \vec v\rangle,
\ee
where $\langle \vec \gamma_0 \vec v, \vec v\rangle=\vec x^\top\vec \gamma_{xx}\vec x+\vec p^\top\vec \gamma_{px}\vec x+\vec x^\top\vec \gamma_{xp}\vec p+\vec p^\top\vec \gamma_{pp}\vec p,$ with $\vec v=\vec x \oplus \vec p\in\rr^\infty$.
Using \eqref{eq:x p expansion} and the linearity of the trace, we can write this as
\begin{align}\label{eq:inner product gamma v selft adjt}
\begin{split}
\langle \vec \gamma_0 \vec v, \vec v\rangle =& \tr[\hat x(f_{\vec x}
)\,\hat x(f_{\vec x})\hat \varrho_0]+\tr[\hat p(f_{\vec p})\,\hat x(f_{\vec x})\hat \varrho_0]\\
&+\tr[\hat x(f_{\vec x})\,\hat p(f_{\vec p})\hat \varrho_0]+\tr[\hat p(f_{\vec p})\,\hat p(f_{\vec p})\hat \varrho_0],
\end{split}
\end{align}
where we have used the notation of Eq. \eqref{eq:x p f} and defined
\be\label{eq:f vec def}
f_{\vec z}=f_{\vec z}(x)=\sum_{k=1}^\infty P_{k-1}(x) z_k, \quad \vec z=\vec x, \vec p.
\ee
Since $\{P_k\}_{k=0}^\infty$ form a complete orthonormal system, for every function $g\in L^2(\mu_0(dx))$ we can associate uniquely (once a sign convention for the set $\{P_k\}_{k=0}^\infty$ has been chosen) a vector $\vec g\in \nn^\infty$. This is achieved by writing $g$ in the form
\be\label{eq:g in terms of its vec}
g(x)=\sum_{k=1}^\infty g_k P_{k-1}(x).
\ee
Let us define the set of functions in $L^2(\mu_0(dx))$ with a specific normalization,
\be
L^2_\textup{nor}(\mu_0(dx),E):=\left\{f\in L^2(\mu_0(dx))\; \Big{|} \int \mu_0(dx)f^2(x)=E \right\}.
\ee
Writing a function $g\in L^2(\mu_0(dx))$ in terms of its associated vector $\vec g$, using Eq. \eqref{eq:g in terms of its vec} we achieve
\be
\int \mu_0(dx) g^2(x)=\sum_{m,n=1}^\infty g_m g_n\int \mu_0(dx)P_{m-1}(x)P_{n-1}(x)=\sum_{n=1}^\infty g_n^2=\|\vec g\|^2.
\ee
Thus a function $g\in L^2(\mu_0(dx))$ is in  $ L^2_\textup{nor}(\mu_0(dx),E)$ iff its associated vector $\vec g$ satisfies $\|\vec g\|^2=E$. Hence noting that $\| \vec v\|^2=\|\vec x\|^2+\|\vec p\|^2$, from Eqs. \eqref{eq:gamma self adjt}, \eqref{eq:inner product gamma v selft adjt}, \eqref{eq:f vec def} we conclude
\begin{align}\label{eq:op nr gamma functional form}
\begin{split}
\|\vec \gamma_0\|=\sup_{\substack{f_1\in L^2_\textup{nor}(\mu_0(dx),E),\\ f_2\in L^2_\textup{nor}(\mu_0(dx),1-E),\\ E\,\in[0,1]}} \quad \Big (
&\tr[\hat x(f_1
)\,\hat x(f_1)\hat \varrho_0]+\tr[\hat p(f_2)\,\hat x(f_1)\hat \varrho_0]\\
&+\tr[\hat x(f_1)\,\hat p(f_2)\hat \varrho_0]+\tr[\hat p(f_2)\,\hat p(f_2)\hat \varrho_0]\Big).
\end{split}
\end{align}
Eq. \eqref{eq:op nr gamma functional form} represents an alternative method to calculate $\|\vec \gamma_0\|$ in which one has to take the supremum over functions $f_1,f_2$.

\subsubsection{Examples for $\| \vec \gamma_0\|$}\label{sec:examples gamma}
Since the operators $\hat x_n$, $\hat p_n$ in Eq. \eqref{eq:x p expansion} only act non trivially on the bath $\mathcal{B}$, one only needs to specify $\hat \varrho_\textup{B}:=\tr_\mathcal{S}[\hat\varrho_0]$ (rather than the full initial quantum state $\hat \varrho_0$ on $\mathcal{S}\otimes\mathcal{B}$ ) in order to calculate $\| \vec \gamma_0\|$. For the interest of finding a simple example, it is useful to write $\hat \varrho_\textup{B}$ the terms of the local number basis of the raising and lowering operators $b^\dag_n(0),b_n(0)$ defined in section \ref{Proofs}.  This basis was first introduced on page 165 of \cite{misc14}, and forms a complete basis for quantum states on $\mathcal{B}$. For every $n\in \nn_0$, its associated number basis is generated by the usual relations $b_n(0)|{0}\rangle_n=0$, $b_n^\dag(0)|{m}\rangle_n=\sqrt{m+1}|{m+1}\rangle_n$. For our example, we will focus on the special case where there are $n_0$ excitation's in each oscillator. The state is then
\be\label{eq:rho B example}
\hat \varrho_\textup{B}=\hat \rho_0\otimes\hat\rho_1\otimes\hat\rho_2\ldots,
\ee
where $\hat \rho_n=|n_0\rangle_n\!\langle n_0|,\,$ $n\in \nn_0$. For $\hat \varrho_\textup{B}$ in  Eq. \eqref{eq:rho B example},\, Eq. \eqref{eq:gamma mat thorem 1} takes the form,
\be\label{eq: gamma matrix example}
\vec \gamma_0=
\left(\begin{array}{cc}
(n_0+1/2)\id & (\mi/2)\id\\
(-\mi/2)\id & (n_0+1/2)\id
\end{array}\right),
\ee
where the $\vec \gamma_0$ has been written in the same block form as in Eq. \eqref{eq:gamma mat thorem 1}. Eq. \eqref{eq: gamma matrix example} has two degenerate eigenvalues, namely $n_0$ and $n_0+1$. Thus
\be\label{eq:op norm gamma matrix example}
\|\vec \gamma_0\|= n_0+1.
\ee
A particular physically transparent case of Eq. \eqref{eq:op norm gamma matrix example}, is when $n_0=0$. For this choice, $\hat\varrho_\textup{B}$ is the vacuum state of the bath Hamiltonian $d\hat{\mathrm{\Gamma}}(G)$. Indeed, as pointed out in \cite{misc14}, the vacuum state of $d\hat{\mathrm{\Gamma}}(G)$ is the same vacuum state as that defined by the number basis of the raising and lowering operators $b^\dag_n(0),b_n(0)$.

\subsection{Alternative expressions for $\|\vec \gamma_0'\|$}\label{sec:Alternative expressions for vec gamma_0'}
In this section we derive similar expressions to those of section \ref{sec:alternative expressions for gamma 0} for $\|\vec \gamma_0'\|$ appearing in theorem \ref{theorem dis 2}. One may wonder whether $\{\hat x_n,\hat p_n\}_{n=1}^\infty$ and $\{\hat x_n',\hat p_n'\}_{n=1}^\infty$ (defined in Eqs. \eqref{eq:x p expansion}, \eqref{eq:x p expansion 2} respectively) are related via a unitary transformation of the form discussed in section \ref{Basis invariance of vec gamma 0} and thus whether $\|\vec \gamma_0'\|$ and $\|\vec \gamma_0\|$ are equal. This turns out not to be the case as we will now discover .
\subsubsection{Relationship between $\vec \gamma_0$ and $\vec \gamma_0'$}
From Eqs. (158) and (161) in \cite{misc14}, one can easily verify using the orthogonality and completeness relations of the underlying orthogonal polynomials that $\vec R'=\vec C \vec{R}$,  $\vec{R}':=(\hat{x}'_1,\hat{x}'_2,\ldots,\hat{p}'_1,\hat{p}'_2,\ldots)^\top$,\\ $\vec R:=(\hat x_1,\hat x_2,\ldots,\hat p_1,\hat p_2,\ldots)^\top,\,$ where
\be
\vec C=
\left(\begin{array}{cc}
\vec A & \vec 0\\
\vec 0 & \vec B
\end{array}\right),
\ee
with
\begin{align}
\begin{split}
[\vec A]_{n,m}&=\sqrt{\frac{2}{\omega_{max}}}\int \mu_0(dx)xP_n'(x^2) P_m(x),\\
[\vec B]_{n,m}&=\sqrt{2\,\omega_{max}}\int \mu_0(dx) P_n'(x^2) P_m(x).
\end{split}
\end{align}
This gives us the relation
\be
\vec \gamma_0'=\vec C \vec \gamma_0\vec C^\top,
\ee
and one can easily verify that $\vec C$ satisfies
\be
\vec C^\top \vec\Omega \vec C=\vec \Omega, \quad \vec \Omega:=
\left(\begin{array}{cc}
\vec 0 & \vec \id \\
-\vec \id  & \vec 0
\end{array}\right).
\ee
However, since $\vec A$ and $\vec B$ are manifestly not unitary, $\vec C$ does not satisfy $\vec C\vec C^\dag=\vec C^\dag \vec C=\id$ and thus $\vec \gamma_0'$, $\vec \gamma_0$ are not unitarily equivalent. Therefore, there is no reason to believe that $\|\vec\gamma_0'\|$ and $\|\vec\gamma_0\|$ are equal in general.
\subsubsection{Alternative expression for $\vec \gamma_0'$}\label{When vec gamma_0' is a normal operator}
Similarly to section \ref{sec:When vec gamma_0 is self-adjoint}, we find that
$\vec \gamma_0'$ is self-adjoint. Therefore
\be\label{eq:gamma self adjt 2}
\|\vec \gamma_0'\|=\sup_{\|\vec v\|=1}\; \langle \vec \gamma_0' \vec v, \vec v\rangle,
\ee
where $\langle \vec \gamma_0' \vec v, \vec v\rangle=\vec x^{\prime\top}\vec \gamma_{xx}'\vec x+\vec p^\top\vec \gamma_{px}'\vec x+\vec x^\top\vec \gamma_{xp}'\vec p+\vec p^\top\vec \gamma_{pp}'\vec p,$ with $\vec v=\vec x \oplus \vec p\in\rr^\infty$.
Using \eqref{eq:x p expansion} and the linearity of the trace, we can write this as
\begin{align}\label{eq:inner product gamma v selft adjt 2}
\begin{split}
\langle \vec \gamma_0' \vec v, \vec v\rangle =&\, \omega_{max}\tr[\hat x(f_{1,\vec x}
)\,\hat x(f_{1,\vec x})\hat \varrho_0]+\tr[\hat p(f_{2,\vec p})\,\hat x(f_{1,\vec x})\hat \varrho_0]\\
&+\tr[\hat x(f_{1,\vec x})\,\hat p(f_{2,\vec p})\hat \varrho_0]+\frac{1}{\omega_{max}}\tr[\hat p(f_{2,\vec p})\,\hat p(f_{2,\vec p})\hat \varrho_0],
\end{split}
\end{align}
where we have used the notation of Eq. \eqref{eq:x p f} and defined
\begin{align}\label{eq:f vec def 2}
f_{1,\vec z}&=f_{1,\vec z}(x)=\sum_{k=1}^\infty P'_{k-1}(x^2) z_k, \\
f_{2,\vec z}&=f_{2,\vec z}(x)=xf_{1,\vec z}(x)
,\quad \vec z=\vec x, \vec p.
\end{align}
Since $\{P_n'\}_{n=0}^\infty$ form a complete orthogonal system, for every function $g\in L^2(\mu_1(dx))$ we can associate uniquely (once a sign convention for the set $\{P_n'\}_{n=0}^\infty$ has been chosen) a vector $\vec g'\in\nn^\infty$. This is achieved by writing $g$ in the form
\be\label{eq:g prime vec}
g(x)=\sum_{n=1}^\infty g'_n P'_{n-1}(x),
\ee
from which it follows from the orthogonality relations
\be\label{eq:g prime vec less infty}
\int \mu_1(dx)g^2(x)=\|\vec g'\|^2 <\infty \quad \textup{iff}\; g\in L^2(\mu_1).
\ee
By defining the sets
\begin{align}\label{def: even odd sets}
\begin{split}
S_\textup{even}&:=\Big\{ f\; \Big{|} \exists\, f'\in L^2(\mu_1)\, s.t. \,f'(x^2)=f(x) \Big\},  \\
S_\textup{odd}&:=\Big\{ f\; \Big{|} \exists\, f'\in L^2(\mu_1)\, s.t. \,xf'(x^2)=f(x) \Big\},
\end{split}
\end{align}
we see that $f_{1,\vec Z}\in S_\textup{even}$ and $f_{2,\vec Z}\in S_\textup{odd}$ iff $\|\vec z\| <\infty$. For every $g\in S_\textup{even}$, and $h\in S_\textup{odd}$ we associate uniquely (once a sign convention for the set $\{P_n'\}_{n=0}^\infty$ has been chosen) a vector $\vec g^e\in\nn^\infty$ and $\vec h^o\in\nn^\infty$ respectively. This is achieved by writing $g$ and $h$ in the form
\be
g(x)=\sum_{n=1}^\infty g^e_n P'_{n-1}(x^2), \quad h(x)=\sum_{n=1}^\infty h^o_n x P'_{n-1}(x^2).
\ee
From Eqs. \eqref{eq:g prime vec}, \eqref{eq:g prime vec less infty} and definitions \eqref{def: even odd sets} it follows that the associated vectors $\vec g^e,$ $\vec h^o$ with every $g\in S_\textup{even}$ and $h\in S_\textup{odd}$ respectively satisfy $\|\vec g^e\|<\infty$ and $\|\vec h^o\|<\infty$.
Note that for every $g\in S_\textup{even}$ its associated vector satisfies
\begin{align}\label{eq:integral even func}
\begin{split}
2\int \mu_0(dx) x g^2(x)&= \sum_{n,m=1}^\infty g^e_n g^e_m 2\int dx \frac{J(x)}{\pi} x P'_n(x^2)P'_m(x^2)\\
&=\sum_{n,m=1}^\infty g^e_n g^e_m\int \mu_1(dy)P'_n(y)P'_m(y)=\sum_{n=0}^\infty \left(g^e_n\right)^2=\|\vec g^e\|^2,
\end{split}
\end{align}
where we used the change of variable $y=x^2$.
With this observation in mind, we define the set
\be
S_\textup{even}(E):=\Big\{ f\in S_\textup{even} \Big{|}\, 2\int \mu_0(dx) x f^2(x)=E \Big\}.
\ee
From Eq. \eqref{eq:integral even func} it follows that $g\in S_\textup{even}$ is in $S_\textup{even}(E)$ iff its associated vector $\vec g^e$ satisfies $\| \vec g^e\|^2=E$.
Similarly to Eq. \eqref{eq:integral even func} we find for every $h\in S_\textup{odd}$,
\be
2\int \mu_0(dx) h^2(x)/x=\|\vec h^o\|^2
\ee
and thus we define the set
\be
S_\textup{odd}(E):=\Big\{ f\in S_\textup{odd} \Big{|}\, 2\int \mu_0(dx) f^2(x)/x=E \Big\},
\ee
finding that $g\in S_\textup{odd}$ is in $S_\textup{odd}(E)$  iff its associated vector $\vec g^o$ satisfies $\| \vec g^o\|^2=E$. Thus noting that $\|\vec v\|^2=\|\vec x\|^2+\| \vec p\|^2$, from Eqs. \eqref{eq:gamma self adjt 2}, \eqref
{eq:inner product gamma v selft adjt 2} it follows
\begin{align}\label{eq:op nr gamma functional form 2}
\begin{split}
\|\vec \gamma_0'\|=\sup_{\substack{f_1\in S_\textup{even}(E),\\ f_2\in S_\textup{odd}(1-E),\\ E\,\in[0,1]}} \quad \Big (
&\omega_{max}\tr[\hat x(f_1
)\,\hat x(f_1)\hat \varrho_0]+\tr[\hat p(f_2)\,\hat x(f_1)\hat \varrho_0]\\
&+\tr[\hat x(f_1)\,\hat p(f_2)\hat \varrho_0]+\frac{1}{\omega_{max}}\tr[\hat p(f_2)\,\hat p(f_2)\hat \varrho_0]\Big).
\end{split}
\end{align}
Eq. \eqref{eq:op nr gamma functional form 2} demonstrates that $\|\vec\gamma_0'\|$ can also be calculated by taking the supremum over functions $f_1,f_2$ in an appropriately defined space.

\subsubsection{Examples for $\| \vec \gamma_0'\|$}\label{sec:examples gamma'} This section will follow very closely the example of section \ref{sec:examples gamma} and will use notation defined there. It is useful to write $\hat \varrho_\textup{B}$ the terms of the local number basis of the raising and lowering operators $b^\dag_n(1),b_n(1)$ defined in section \ref{Proofs}. This basis was first introduced on page 165 of \cite{misc14}, and forms a complete basis for quantum states on $\mathcal{B}$. For every $n\in \nn_0$, its associated number basis is generated by the usual relation $b_n(1)|{0}\rangle_n=0$, $b_n^\dag(1)|{m}\rangle'_n=\sqrt{m+1}|{m+1}\rangle'_n$. Let us define the state
\be\label{eq:rho B example'}
\hat \varrho_\textup{B}=\hat \rho_0'\otimes\hat\rho_1'\otimes\hat\rho_2'\ldots,
\ee
where $\hat \rho_n'=|n_0\rangle_n'\!\langle n_0|',\,$ $n\in \nn_0$. For $\hat \varrho_\textup{B}$ in  Eq. \eqref{eq:rho B example'},\, Eq. \eqref{eq:gamma mat thorem 2} takes the form,
\be\label{eq: gamma matrix example'}
\vec \gamma_0'=
\left(\begin{array}{cc}
(n_0+1/2)\id & (\mi/2)\id\\
(-\mi/2)\id & (n_0+1/2)\id
\end{array}\right).
\ee
As noted in section \ref{sec:examples gamma}, the two degenerate eigenvalues of Eq. \eqref{eq: gamma matrix example'} are $n_0$ and $n_0+1$ and we thus find
\be\label{eq:op norm gamma matrix example'}
\|\vec \gamma_0'\|= n_0+1.
\ee
We note that although the r.h.s. of Eqs. \eqref{eq:op norm gamma matrix example} and \eqref{eq:op norm gamma matrix example'} are the same, the examples are very different since the states defined in Eqs. \eqref{eq:rho B example} and \eqref{eq:rho B example'} are different states. This is because they are defined in different basis. For example, the state in Eq. \eqref{eq:rho B example'} for $n_0=0$ is \textit{not} the vacuum state of $d\hat{\mathrm{\Gamma}}(G)$.

\end{document}